\documentclass{ws-procs9x6-cpt22}
\begin{document}

\newcommand{\refeq}[1]{(\ref{#1})}
\def\etal {{\it et al.}}

\title{Recent experimental progress on probing Lorentz Violation in pure gravity for $d\!=\!6$ }

\author{Tao Jin, Jia-Rui Li, Yu-Jie Tan, and Cheng-Gang Shao}
\address{MOE Key Laboratory of Fundamental Physical Quantities Measurements $\&$ Hubei Key \\Laboratory of Gravitation and Quantum Physics,\\PGMF and School of Physics, Huazhong University of Science and Technology,\\Wuhan 430074, Peoples Republic of China}
\begin{abstract}
Short-range experiment with striped design of the test and source masses offers uniquely sensitive probes for Lorentz Violation. In our previous work, we proposed to combine the horizontally and vertically striped experiments to constrain the violating parameters. Here, we further point out adopting the vertically striped structure with the experimental setup placed at two azimuth angles, which has more experimental operability. For the experiment, we have completed the assembly and measurement of the torsion pendulum.
\end{abstract}
\bodymatter
\section{Introduction}
Lorentz symmetry, the idea that physical laws are unchanged under rotations and boosts, is the cornerstone of both General Relativity and the Standard Model. In SME frame, the pure gravity Lorentz Violation (LV) effects cause corrections to the Newton potential, which are related to the violating coefficients\cite{1}. As the corrected potentials for $d\!\geq{}\!6$ are inversely proportional to power of distance, the short-range gravity experiments are sensitive to test these LV effects. By analyzing the experimental data of testing the gravitational inverse square law, our group collaborating with Indiana University has given the best LV constraints for $d=6$ \cite{2}. Moreover, to obtain a better constraint, we designed a special experimental scheme for LV test, in which both the test and source masses are in striped structure to enhance the LV signal\cite{3,4}, and the experiment is under going. In this work, we mainly introduce the experimental progress and the related theoretical analysis.
\section{Experimental scheme of LV test}
\begin{figure}[!t]
\begin{center}
\includegraphics[width=3.0in]{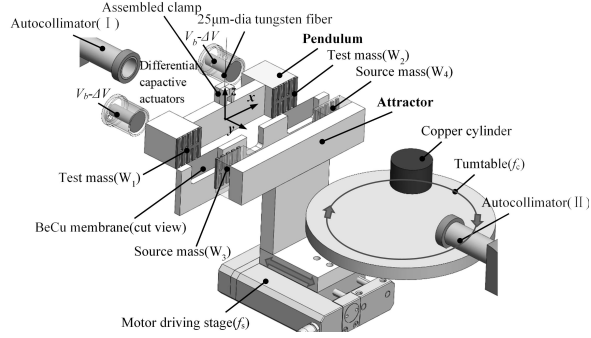}
\end{center}
\caption{Schematic diagram of the LV experimental setup (not to scale). The pendulum twist is measured by an autocollimator, and controlled by two differential capacitive actuators. The rotating copper cylinder is set to calibrate the sensitivity of the pendulum.}
\label{j.fig1}
\end{figure}

 The experimental principle has been described in the previous work\cite{3,4}, and the experimental apparatus is similar to the inverse-square experiment HUST-2012, shown in Fig.\ref{j.fig1}. The LV force between test and source masses varies with earth's rotation by the sidereal time. Meanwhile, we introduce another modulation by changing the distance between the test and source masses at frequency $\omega_s$ for reducing the torque noise.
The measured LV torque in the experiment can be expressed as:
\begin{equation}
\tau _\textup{measured}^z(T) =\tau _\textup{LV}(T)\cos (\omega_sT+\varphi)
\label{j:eq3},
\end{equation}
with
\begin{equation}
\tau _\textup{LV}(T) =C_0+\sum_{m=1}^4{C_m\cos (m \omega_\oplus T)+S_m\sin (m\omega_\oplus T)}
\label{j:eq4}.
\end{equation}
Here, $\omega_\oplus \approx 2\pi /{}$(23 h 56 min) is the Earth's sidereal frequency, and $\varphi$ is initial phase. The nine Fourier amplitudes $C_0, C_m, S_m$ can be expressed in terms of the LV coefficients $k_{jm}^{N(6)}$ and the transfer coefficients $\Gamma_j$, shown in Table.\ref{j.tbl1}. $\Gamma_j$ is related to the geometries of the test and source masses, the colatitude $\chi$, and the angle $\theta$ between the $x$ axis and local south. To pursuit the good constraints, any experimental design should make $\Gamma_j$ as large as possible. According to the typical design parameters, we numerically calculated the transfer functions for the test and source masses in horizontal stripes, as well as vertical case. The result shows $\theta=3\pi/5$ and $\pi$/6 in the vertically striped case are the good choice for the experiment (the corresponding values of the transfer coefficients are listed in Table.\ref{j.tbl1}, which is more sensitive than that of HUST-2012 almost one order of magnitude). Thus, we can use one experimental setup to construct two experiments by setting the angle at these two values, respectively. Combining the experimental data, the 14 LV parameters can be constrained independently.

\begin{table}
\tbl{Transfer coefficients $\Gamma_j$ for the vertically striped experiment.}
{\begin{tabular}{@{}cccccc@{}}\toprule

Quantity &Expression &Transfer &HUST-2012 &\multicolumn{2}{c}{New experimental design}\\ & &coefficients & &$\theta =3 \pi /5$ &$\theta =\pi /6$\\

\colrule
$C_0$ & $\Gamma_1k_{2,0}+\Gamma_2k_{4,0}$                              & $\Gamma_1$    & -0.9 & -25.3 & -10.7 \\
      &                                                                & $\Gamma_2$    & 0.0 & 19.0 & -8.2  \\
\colrule
$C_2$ & $ \Gamma_3\textup{Re}k_{2,2}+\Gamma_4\textup{Im}k_{2,2}$       & $\Gamma_3$    & -0.4 & 17.8 &12.9 \\
      & $+\Gamma_5\textup{Re}k_{4,2}+\Gamma_6\textup{Im}k_{4,2}$       & $\Gamma_4$    & 0.3 & -14.5  &36.4 \\
$S_2$ & $ \Gamma_4\textup{Re}k_{2,2}-\Gamma_3\textup{Im}k_{2,2}$       & $\Gamma_5$    &-0.9 & -14.1 & -14.7 \\
      & $+\Gamma_6\textup{Re}k_{4,2}-\Gamma_5\textup{Im}k_{4,2}$       & $\Gamma_6$    & -0.4 & 4.1 & 4.9 \\
\colrule
$C_4$ & $ \Gamma_7\textup{Re}k_{4,4}+\Gamma_8\textup{Im}k_{4,4}$       & $\Gamma_7$    & -3.7 & 1.5 & 22.9 \\
$S_4$ & $ \Gamma_8\textup{Re}k_{4,4}-\Gamma_7\textup{Im}k_{4,4}$       & $\Gamma_8$    & -0.4 & 14.3 & 33.1 \\
\colrule
$C_1$ & $ \Gamma_9\textup{Re}k_{2,2}+\Gamma_{10}\textup{Im}k_{2,2}$    & $\Gamma_9$    & -1.6 & 20.5 & 13.3 \\
      & $+\Gamma_{11}\textup{Re}k_{4,2}+\Gamma_{12}\textup{Im}k_{4,2}$ & $\Gamma_{10}$ & -0.5 & -4.0 & 18.7 \\
$S_1$ & $ \Gamma_{10}\textup{Re}k_{2,2}-\Gamma_9\textup{Im}k_{2,2}$    & $\Gamma_{11}$ & 1.0 & 18.9 &8.5 \\
      & $+\Gamma_{12}\textup{Re}k_{4,2}-\Gamma_{11}\textup{Im}k_{4,2}$ & $\Gamma_{12}$ & -0.4 &  -19.4&2.9    \\
\colrule
$C_3$ & $ \Gamma_{13}\textup{Re}k_{4,4}+\Gamma_{14}\textup{Im}k_{4,4}$ & $\Gamma_{13}$ & -0.1 &-6.4  &16.8  \\
$S_3$ & $ \Gamma_{14}\textup{Re}k_{4,4}-\Gamma_{13}\textup{Im}k_{4,4}$ & $\Gamma_{14}$ &  -0.4&16.7  &-22.6  \\
\botrule
\end{tabular}
}
\label{j.tbl1}
\end{table}
\section{Error analysis and experimental progress}
The LV coefficients $k_{jm}^{N(6)}$ can be constrained by a fit of the Fourier amplitudes $C_0, C_m, S_m$ data. Thus, the constrained levels of LV coefficients are determined by the uncertainties of the Fourier amplitudes. The value of the dc component $C_0$ extracted from recorded data $\tau _\textup{measured}^z(T)$ is restricted by the systematic errors arising from uncertainties in the dimensions and locations of the test and source masses. The sidereal-harmonic amplitudes $C_m, S_m$ are extracted from $\tau _\textup{LV}(T)$ demodulated from the recorded data $\tau _\textup{measured}^z(T)$, and their uncertainties are dominated by the statistical uncertainty which is at the same level $0.4 \times 10^{-16}$ Nm for each harmonic.
\begin{figure}
\begin{center}
\includegraphics[width =0.42\textwidth]{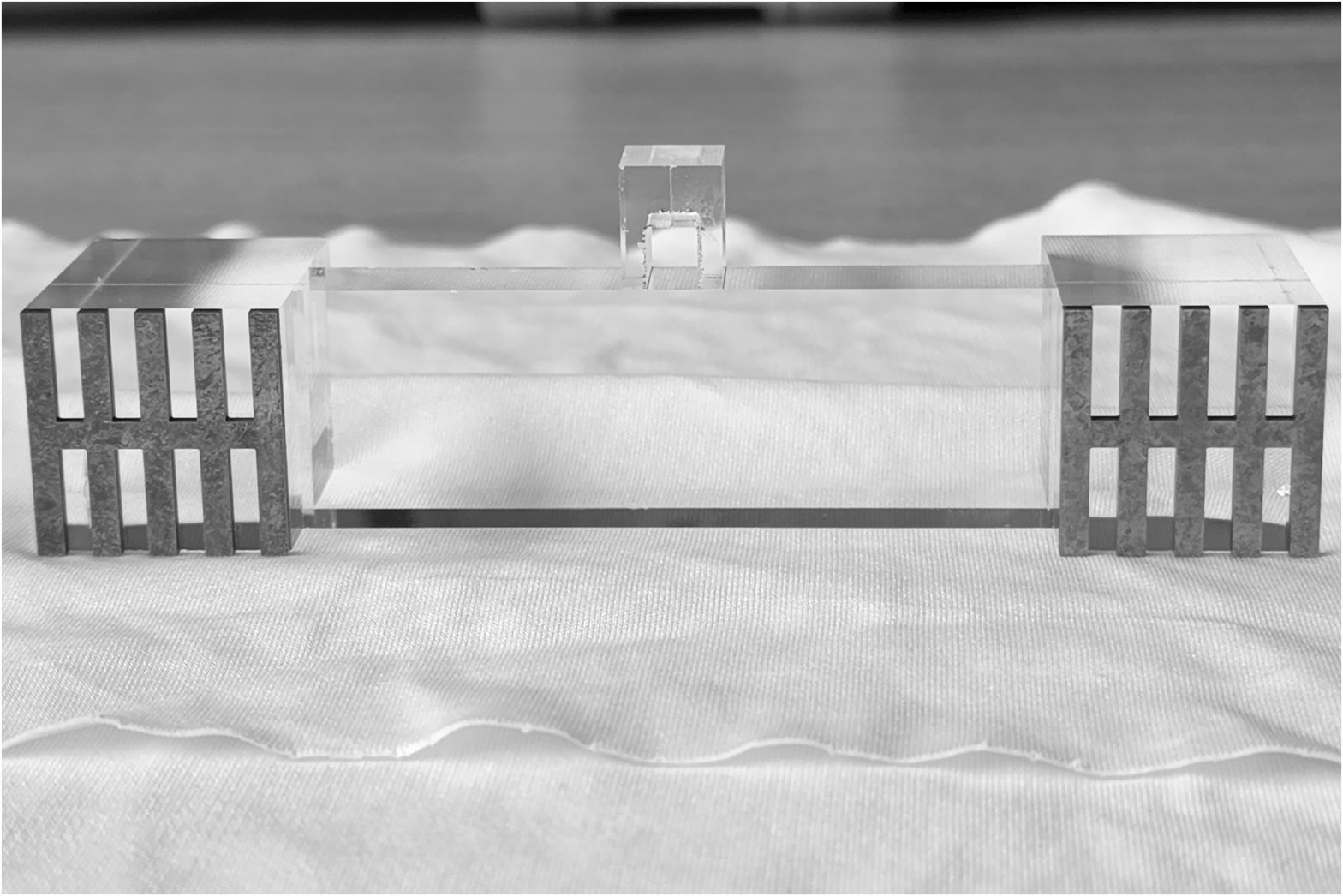}
\includegraphics[width = .225\textwidth]{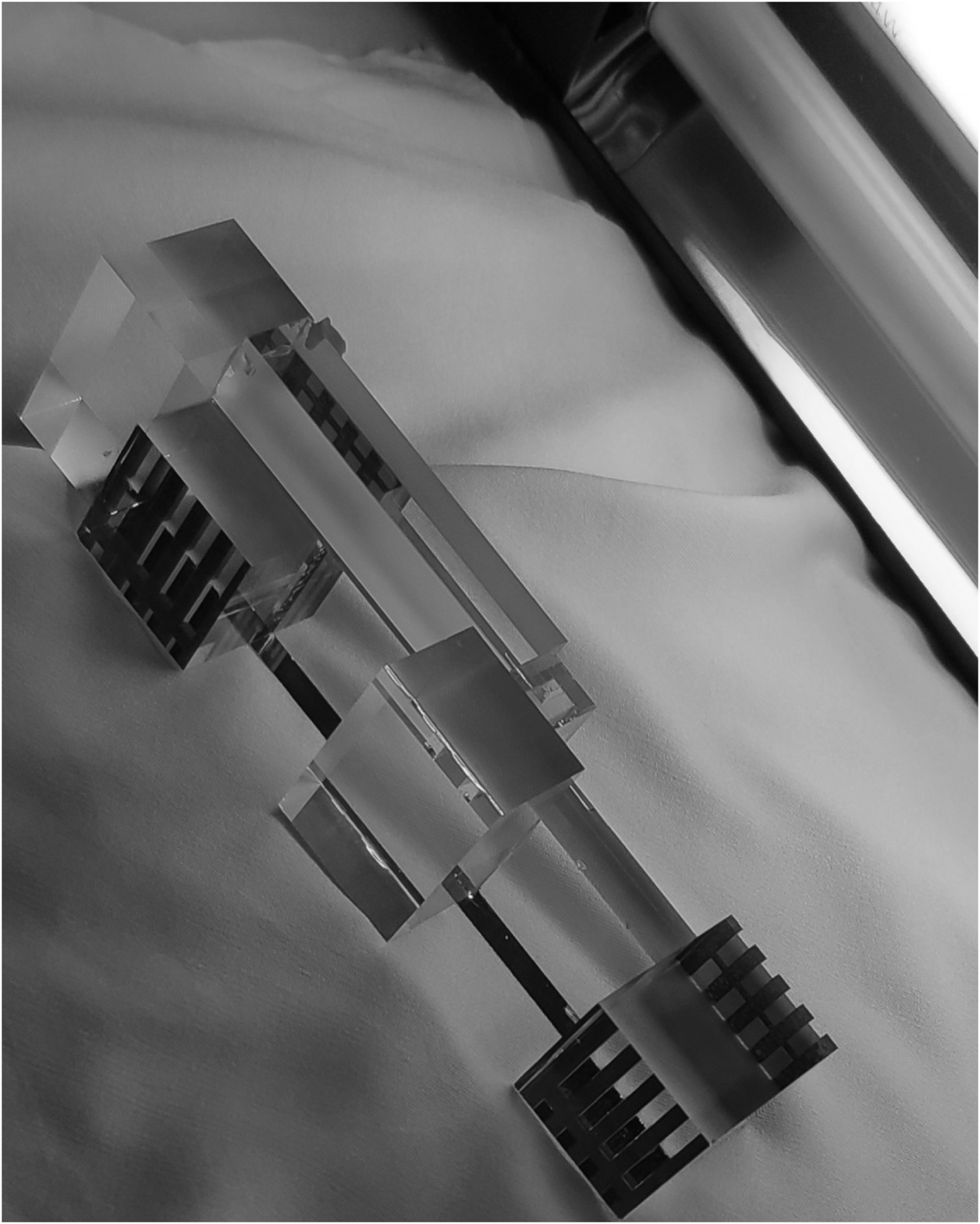}
\includegraphics[width = .225\textwidth]{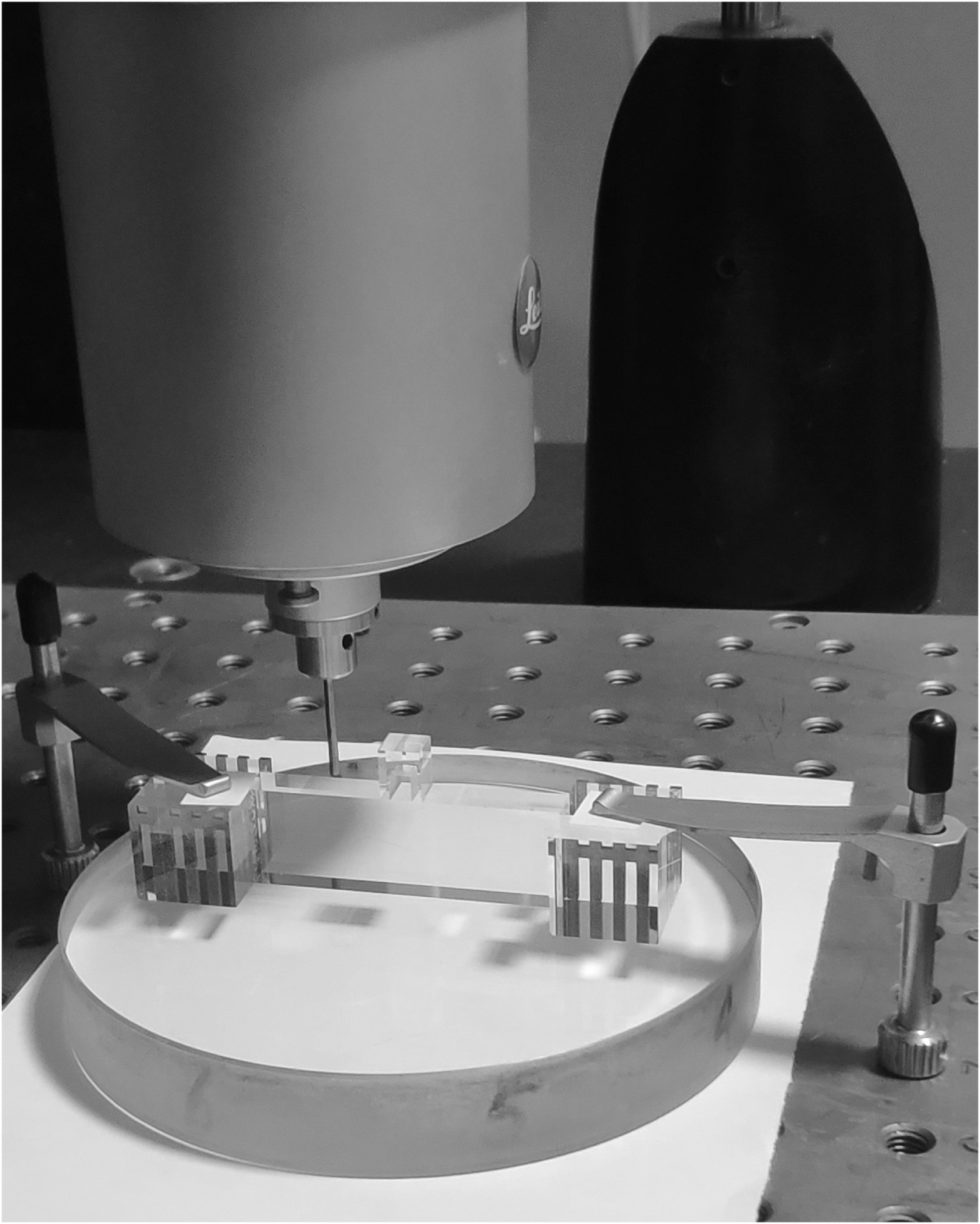}
\end{center}
\caption{Assembling and measurement of pendulum.}
\label{j.fig2}
\end{figure}

At present, the experimental components, such as striped tungsten slices and glass blocks, have been machined; the geometry parameters of the tungsten slices and some core parts have been measured in a double-blind way, and the result shows the machining accuracy meets the experimental requirements; the torsion pendulum has been assembled and measured, shown in Fig.\ref{j.fig2}. Next, we will suspend the pendulum to test the background, assemble the source mass and put it into the vacuum chamber, and finally record the experimental data after everything is normal.
\section{Conclusion}
In this work, we mainly review and slightly improve the experimental scheme we proposed to test LV in the previous work, which can improve the current constraints on the LV coefficients by more than one order of magnitude. Currently, the experiment is proceeding as planned, and we hope the experimental results can be obtained within two years.
\section*{Acknowledgments}
This work was supported by the National Natural Science Foundation of China (Grant Nos. 12175076 and 11925503).

\end{document}